\documentclass[12pt]{article}
\usepackage{graphicx}
\usepackage{epsf}
\newcommand{\be}[1]{\begin{equation}\label{#1}}
\newcommand{\ee}{\end{equation}}
\newcommand{\bea}[1]{\begin{eqnarray}\label{#1}}
\newcommand{\eea}{\end{eqnarray}}
\newcommand{\nn}{\nonumber\\}
\newcommand{\gl}[1]{Eq.\,(\ref{#1})}
\newcommand{\abb}[1]{Fig.\,\ref{#1}}
\newcommand{\tab}[1]{Tab.\,\ref{#1}}
\newcommand{\kap}[1]{Sec.\,\ref{#1}}
\newcommand{\eps}{\varepsilon}

\begin{document}
\vspace*{2cm}
\begin{center}
{\LARGE \bf 
Recycling probability and dynamical\\[2mm]
properties of germinal center reactions}\\
\vspace{1cm}
Michael Meyer-Hermann$^1$, 
Andreas Deutsch$^2$
and
Michal Or-Guil$^2$\\
\vspace{1cm}
$^1$Institut f\"ur Theoretische Physik, TU Dresden,
D-01062 Dresden, Germany\\
E-Mail: meyer-hermann@physik.tu-dresden.de\\
\vspace{2em}
$^2$Max-Planck-Institut f\"ur Physik komplexer Systeme,\\
N\"othnitzer Str.38, D-01187 Dresden, Germany
\end{center}

\vspace{2cm}
\noindent{\bf Abstract:}
We introduce a new model for the dynamics of
centroblasts and centrocytes in a germinal center. 
The model reduces the germinal center reaction to the elements
considered as essential and
embeds proliferation of centroblasts, point mutations of the
corresponding antibody types 
represented in a shape space, differentiation to centrocytes,
selection with respect to initial antigens, differentiation of positively
selected centrocytes to plasma or memory cells and recycling of
centrocytes to centroblasts. We use exclusively
parameters with a direct biological interpretation such that,
once determined by experimental data, 
the model gains predictive power. 
Based on the experiment
of Han et al.\,(1995b) we predict that a high rate of recycling 
of centrocytes to centroblasts is necessary for the 
germinal center reaction to work reliably. 
Furthermore, we find a delayed start of the
production of plasma and memory cells with respect to the start
of point mutations, which turns out to be necessary for 
the optimization process during the germinal center reaction.
The dependence of the germinal center 
reaction on the recycling probability is analyzed.
\vspace*{\fill}
\eject
\newpage

\section{Introduction}

Germinal centers (GCs) are essential
elements of the humoral immune response
(for a review see MacLennan, 1994). In a primary
immunization some B-cells respond to the presented antigen
and are activated. These active cells migrate to follicular
dendritic cells (FDCs) present in lymphoid tissues. 
Unprocessed fragments of the antigen, so called epitopes, 
are expressed on the FDCs.
In the milieu of the FDCs the B-cells are in a phase of
intense proliferation and show the essential features of
B-blasts such that they are denoted as centroblasts. These
proliferating centroblasts together with the FDCs develop
to a GC. 

This morphological development is
initiated through an unidentified differentiation signal
after about three days. The resulting GC is
characterized by two zones (Nossal, 1991): 
the dark zone filled with 
continuously proliferating and mutating centroblasts 
and the light zone
containing the FDCs and centrocytes. The latter are
generated from the centroblasts (Liu et al., 1991), 
which are available in great diversity (Weigert et al., 1970)
because of  a very high 
mutation rate (Nossal, 1991; Berek \& Milstein, 1987).
In contrast to
centroblasts the centrocytes express antibodies 
of cell specific types on their surface.
In this way
they are able to undergo a specific selection process 
through the interaction with the epitopes presented on the
FDCs in the light zone. The diversity of the proliferating
and mutating centroblasts together with the specificity of 
the selection process in the light zone allow an 
optimization of the affinity between the antibodies
and the epitopes. This affinity maturation in the GC leads to a
dominant population of cells with high affinity of the
corresponding antibodies to the epitopes within 7-8 days
after immunization (Jacob et al., 1993). 

Centrocytes undergo
apoptosis if they are not positively selected in an appropriate 
range of time. They are
rescued from apoptosis (Liu et al., 1989; Lindhout et al., 1993;
Lindhout et al., 1995; Choe \& Choi, 1998)
if they are able to bind to the
antigen on the FDCs {\it and} to interact successfully
with T-helper cells present in the external
region of the light zone. Even antigen independent signals
may be required for the centrocytes to survive (Fischer et al., 1998).
Successfully selected centrocytes
receive a differentiation signal which determines their
fate. They may differentiate into memory cells that are
important in the case of a second immunization.
Alternatively, they differentiate to antibody producing
plasma cells. These are intensifying the immune response
in two ways: the number of antibodies increases and their
specificity with respect to the direct unspecific
immune response without GC reaction is optimized.
In both cases the corresponding cell is leaving the
GC environment. 

A third possibility exists for the
positively selected centrocytes: they may be
recycled to centroblasts and reenter the highly proliferating
and mutating stage of development in the dark zone 
(Kepler \& Perelson, 1993). 
This recycling hypothesis has been neither directly checked
experimentally (for an indirect check see Han et al., 1995b)
nor -- if true -- the probability 
of recycling is known. It would be of great interest to
know about the necessity of a centrocyte recycling
process for the affinity maturation. The
most evident argument to support the recycling hypothesis is
based on the occurrence of multi-step somatic hypermutations
in the GC reaction. Under the assumption of random
hypermutations, recycling of already mutated cells to
fast proliferating centroblasts appears to be a necessary
process.

In this paper we estimate how large the
recycling probability should be, i.e.\,which proportion of
the positively selected centrocytes
should be recycled to get a reliable GC reaction.
For this purpose we introduce a simplified model of the
GC dynamics based on a functional view on it.
The model differs from
others (Oprea \& Perelson, 1996; Oprea \& Perelson, 1997;
Rundell et al., 1998) by omitting some
details of the GC reaction 
which we consider as unnecessary to understand
the obligatory occurrence of the recycling procedure. 
On the other hand compared to the model introduced in
(Kesmir \& de Boer, 1999) we added a formalism to treat
somatic hypermutation in a shape space, which seems
to be essential to understand the process of affinity
maturation.
A very important difference to related models is that
we embed the experiment (Han et al., 1995b) for the first time.
This experiment 
provides an indirect evidence for the existence of the recycling
process and we will interpret it in a quantitative way in order to
extract information on the recycling probability from it.
We use exclusively parameters with a direct biological
interpretation to ensure the predictive
power of the model. 

In the next section we will define the shape space, in which the
cell population dynamics takes place, and the corresponding
dynamical quantities and parameters 
including their dynamical interdependence.
All parameters introduced in the model
are determined by experimental observations.
The third section is dedicated to the presentation and
interpretation of the model
in comparison with experimental GC reactions. 
We will analyze the stability of our results to verify the
notion of a prediction. Finally, we investigate the dependence
of some characteristics of the germinal center reaction on the
variation of the recycling probability and give some concluding
remarks.

\section{Definition of the model}

\subsection{Shape space and affinity space}

\paragraph{Postulates}
Before defining a model for the GC dynamics it is
necessary to specify the space in which the dynamics takes
place. As we want to represent cells corresponding to
different types of antibodies we use the well known
shape space concept (Perelson \& Oster, 1979).
This ansatz is based on the following assumptions:
\begin{itemize}
\item Antibodies can be represented in a phenotypical shape space.
\item Mutation can be represented in the shape space by next-neighbor jumps.
\item Complementarity of antibodies and antigen.
\item Homogeneity of the affinity weight function on the shape space.
\end{itemize}

\paragraph{Representation of antibody phenotype}
The shape space is taken as a $D$-dimensional finite size 
lattice of discrete equidistant points,
each of them representing one specific antibody type.
Thus the shape space becomes a phenotype space, i.e.\,it is
not primarily a representation of genetic codes but of the
resulting features due to specific genetic codes.

\paragraph{Representation of mutation}
Nevertheless, we want to represent point mutations defined
in an unknown genotype space in the shape space. To this
end we are compelled to formulate the action of a point
mutation in the shape space:\\
{\it A point mutation of a given antibody is represented
in the phenotypical shape space by a jump to one of the
nearest neighbor points.}\\
This assumption is not very spectacular as it only requires
that a point mutation does not lead to a random 
jump of the antibody phenotype, i.e.\,that conformation, electrical properties,
etc. are not dramatically changed. Surely, the mutation of 
some key base pairs may exist which imply fundamental changes
of the encoded antibody features. However, 
these exceptions will not alter the dynamics of the GC
until the number of such key base pairs remains small with respect
to the number of {\it smooth} mutation points.

\paragraph{Antigen representation and complementarity}
A central quantity is the affinity of a given antibody and an antigen epitope.
Having a representation of antibodies in the shape space, a counterpart 
for antigens is necessary. 
We emphasize that the number of possible antibodies is finite, whereas the
diversity of antigens is principally unbounded. Therefore, 
we represent an antigen on the B-cell shape space at the position of
the B-cell with highest affinity to the antigen, i.e.\,at the position of
an antibody with optimal complementarity to the antigen. 

\paragraph{Property of smooth affinity}
Starting from these assumptions (phenotype shape space, 
representation of mutation by next-neighbor jumps, and
complementarity) we have reached a representation
of antibodies and antigens in a shape space,
which has the property of {\it affinity neighborhood}, i.e.\,that
neighbor points in the shape space have comparable affinity to
a given antigen. This property is a direct consequence of the
definition of mutations in a (phenotype) shape space. 
Starting from a 
non-directed mutation as base for affinity maturation in GCs,
we estimate the affinity neighborhood as a necessary
property of the underlying shape space. An initial B-cell with a
particular affinity to a certain antigen must have the possibility of
successively optimizing the affinity, i.e.\,of stepwise climbing
an {\it affinity hill}. If no affinity neighborhood would exist, 
i.e.\,if there was a random affinity distribution on the shape space,
each mutation would lead to a random jump in the affinity.
An optimization of the affinity to the antigen may occur
accidentally in this scenario, but should be a very rare event.

\paragraph{Homogeneous affinity weight function}
Affinity neighborhood allows for the definition of an
{\it affinity weight function} which determines the affinity
of antigen and antibody in dependence of their distance 
in the shape space. We assume this weight
function $a(\phi,\phi^*)$
to apply equally well in all regions
of the shape space which corresponds to a homogeneous
affinity distribution over the shape space, and to be of
exponential form:
\be{affinity}
a(\phi,\phi^*) \;=\; \exp\left(-\frac{||\phi-\phi^*||^\eta}{\Gamma^\eta}\right)
\ee
where $\Gamma$ is the width of the affinity weight function
and $||\cdot ||$ denotes the Euclidean metric.
The question remains with which exponent $\eta$ the distance enters
the exponential weight function and it will be argued that $\eta=2$ is
a reasonable value (see \abb{Gamma}).

\subsection{Formulation of the dynamics}
\label{dynamics}

Having a shape space at hand it is possible to define distributions
$B(\phi)$, $C(\phi)$, $A(\phi)$, and $O(\phi)$
of centroblasts, of centrocytes,
of the presented antigen epitopes, and of the
plasma and memory cells on the shape space
(see Tab.~\ref{variables}).
We focus on the centroblast distribution $B$
and analyze its dynamical behavior in the different phases
of the GC reaction.
\begin{table}[ht]
\begin{center}
\begin{tabular}{|l|c|l|} \hline
Cell type & Variable & Processes \\ \hline
antigen & $A(\phi)$ & interaction with centrocytes \\
centroblasts & $B(\phi)$ & proliferation\\
&& mutation\\
&& differentiation to centrocytes\\
centrocytes & $C(\phi)$ & selection and apoptosis\\
&& recycling to centroblasts\\
&& differentiation to plasma and memory cells\\
plasma and memory cells & $O(\phi)$ & removed from GC\\
\hline
\end{tabular}
\caption[]{\sf The variables in the mathematical model of 
the GC reaction and the processes that are considered.}
\label{variables}
\end{center}
\end{table}
\begin{figure}[ht]
\includegraphics[height=15cm,angle=-90]{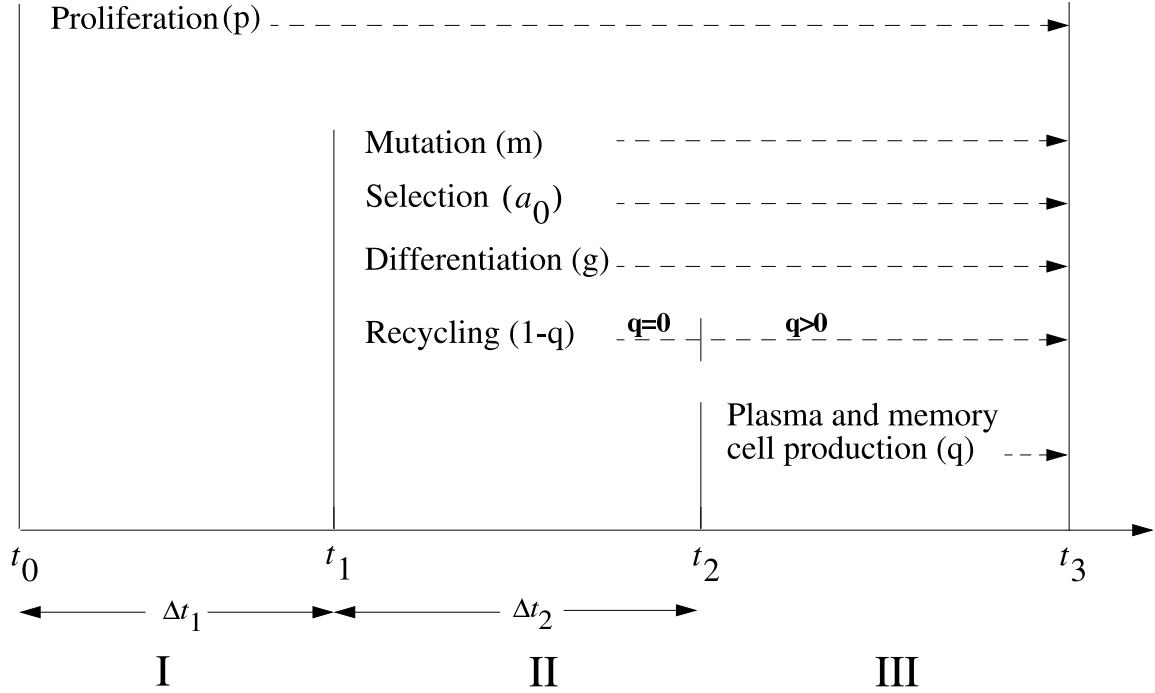}
\caption[]{\sf 
The three model phases of the germinal center reaction: 
(I) proliferation phase (starting at $t_0=-3\,d$), 
(II) optimization phase (starting at $t_1=0\,d$, the duration
$\Delta t_2$ is determined in Sec.~\ref{outputdyn}), 
(III) output phase (ending at $t_3=18\,d$). 
The parameters in brackets
refer to the model equations \gl{pmodell}, 
\gl{modell}, \gl{centrocytes}, and \gl{output}
and can be determined from experimental data 
(see Sec.~\ref{dynamics}-\ref{outputdyn}).}
\label{phases}
\end{figure}

There were several attempts to divide the germinal center
reaction in different working phases. From the
point of view of our model we are led to a new functional
phase distinction of the GC reaction (see Fig.\,\ref{phases}).
Our three-phase description of the GC reaction is
in accordance with most of the models established so far
(see e.g.\,MacLennan, 1994; Liu et al., 1991; Kelsoe, 1996) but
is in contradiction to the time scales found in 
(Camacho et al., 1998), which differ strongly.
It is not intended to find mechanisms of transition between the phases 
of the germinal center reaction. These dynamical phases are
assumed according to experimental observations.

The first phase is the already mentioned phase of fast
proliferation of a low number of seeder cells in the
environment of FDCs. In this phase it is likely that 
somatic hypermutation is not taking place to a relevant
amount (Han et al., 1995b; Jacob et al., 1993; 
McHeyzer-Williams et al., 1993; Pascual et al., 1994b)
such that it may be understood as an enlargement phase 
of the cell pool
available for the following process of affinity maturation.
The corresponding dynamical behavior of the centroblast
distribution is described as
\be{pmodell}
\frac{dB}{dt}(\phi) \;=\; p B(\phi) 
\quad {\rm for} \quad t-t_0<\Delta t_1
\quad,
\ee
where $p$ denotes the proliferation rate and $t_0$ is the
time of first immunization.
This phase lasts $\Delta t_1=3$ days and is well 
established by experiment (Liu et al., 1991).

After three days the optimization phase starts. 
The GC gets its morphological form, i.e.\,dark and light zones emerge.
In the dark zone centroblasts continue to proliferate but,
additionally, somatic hypermutations are broadening
the initial centroblast distribution on the shape space.
At the same time the selection process operates
in the light zone and the selected centrocytes are either recycled
to centroblasts or differentiate to plasma and memory cells.
The development of the centroblast distribution
on the shape space is now described by
\bea{modell}
\frac{dB}{dt}(\phi) &=&
(p - 2pm -g) B(\phi) 
+ \frac{pm}{D} \sum_{||\Delta\phi||=1} B(\phi+\Delta\phi)
\nn
&& + (1-q) \sum_{\phi^*} C(\phi) A(\phi^*) \,a_0 \,a(\phi,\phi^*)
\quad,
\eea
for $t-t_0\ge\Delta t_1$.
Here, $g$ is the rate of centroblast differentiation
into centrocytes. $m$ is the mutation probability, 
$a_0$ is the probability of an optimal
centrocyte to be positively selected,
$q$ is the probability
of differentiation into plasma and memory cells 
for positively selected centrocytes, 
and $D$ is the dimension of the underlying shape space.
The different contributions to \gl{modell} are
discussed in detail below.
Note that this equation does not include eventual finite size
effects due to small populations.

It is not clear a priori, if differentiation into antibody producing
plasma and memory cells is triggered already in this
second phase. To allow a start of the output cell production
delayed by the time interval $\Delta t_2$
we divide the optimization phase into two sub-phases
which differ in the output probability $q$:
\bea{phase2}
q&=&0 \quad {\rm for} \quad 0 \le t-\Delta t_1 < \Delta t_2 \nn
q&>&0 \quad {\rm for} \quad 0 \le t-\Delta t_2 < \Delta t_3
\quad.
\eea
The time delay $\Delta t_2$ will be fixed by experimental
data (see \kap{outputdyn}).
The output phase include optimization and production of
plasma and memory cells and lasts for the remaining
GC life-time, which is about
$\Delta t_3=21{\rm days}-3{\rm days}-\Delta t_2$
(Choe \& Choi, 1998; Kelsoe, 1996; Jacob et al., 1991).
Then, only a few proliferating B-blasts remain in the
environment of the FDCs (Liu et al., 1991).

\subsubsection*{Proliferation}

During the whole GC reaction a fast proliferation
of B-cells takes place. After the activation of B-cells by an
interaction with an antigen, these move to FDCs and undergo
a fast proliferation phase in their environment 
(Liu et al., 1991; Hanna, 1964; Zhang et al., 1988; Grouard et al., 1995). 
It is likely that the fast proliferation is triggered by the
dendritic cells (Dubois et al., 1999).
After about three days (Liu et al., 1991)
when the GC starts to develop its characteristic light and
dark zone (Camacho et al., 1998),
a fast proliferation of centroblasts is still observed in the dark zone.
The extremely high rate of 
proliferation could be determined 
to be (Liu et al., 1991; Zhang et al., 1988)
\be{p}
\frac{p}{\ln(2)}\;\approx\; \frac{1}{6\,h}
\quad,
\ee 
a result known already for a long time (Hanna, 1964). 
The population of centroblasts at $\phi$ grows exponentially, 
which is represented
by the term $pB(\phi)$ in \gl{pmodell} and \gl{modell}.

\subsubsection*{Mutation}

It is likely that somatic hypermutation does not occur in the
proliferation phase of the GC reaction 
(Han et al., 1995b; Jacob et al., 1993; 
McHeyzer-Williams et al., 1993; Pascual et al., 1994b).
On the other hand the growth of the centroblast population is reduced 
by possible mutations from state $\phi$ to its neighbors
$\phi+\Delta\phi$, where $||\Delta\phi||=1$. 
In a continuous space this corresponds to a diffusion process
as used in (Perelson \& Wiegel, 1999) to represent mutation.
Each pair
of cells produced by proliferation will populate a neighbor
state with a probability of $m$. This mutation probability turns out
to take extremely high values of $m\approx 1/2$ 
(Nossal, 1991; Berek \& Milstein, 1987),
which corresponds
to a factor $10^7$ larger probability compared to mutations outside
the GCs (Janeway \& Travers, 1997). We want to point out that here only
point mutations of phenotypical relevance are taken into account.
As a consequence the population
growth by proliferation is reduced by the important
amount $2pm$, which results in an effective proliferation rate 
at $\phi$ of $p(1-2m)$.
One observes that for $m=1/2$ the two first terms in \gl{modell}
cancel. This corresponds to the situation that in each centroblast
replication one new cell remains in the old state and the second new cell
mutates to a neighbor state such that the total number of cells in
state $\phi$ remains unaltered.

In the same way as mutation depopulates the state $\phi$ in the 
shape space and populates its neighbors, $\phi$ is populated by
its neighbors. This gives rise to the non-diagonal term in \gl{modell}.
Each neighbor of $\phi$ mutates with the same rate $2pm$ from 
$\phi+\Delta\phi$ to one of its $2D$ neighbors. 
Only the mutation from $\phi+\Delta\phi$ to the particular neighbor
$\phi$ enhances the population of centroblasts at $\phi$
such that the rate is weighted by the inverse number of neighbors
$1/(2D)$ and is summed over all possible neighbors of $\phi$.

\subsubsection*{Selection, recycling and cell production}

After the establishment of the light and dark zones in
the GC the differentiation of centroblasts to
antibody presenting centrocytes is triggered to allow
a selection process in the light zone. The centroblasts
in the shape space state $\phi$ are diminished 
with the rate of differentiation $g$, 
leading to the term $-gB(\phi)$ in \gl{modell}.
The centrocyte population $C(\phi)$ of type $\phi$ is enhanced
simultaneously by the same amount of cells:
\be{centrocytes}
C(\phi) \;=\; +gB(\phi)
\quad.
\ee
The centrocytes move to the light zone where they
undergo a selection process. 
Their further development is splitted
three-fold. The non-selected centrocytes die through apoptosis
and are eliminated from the GC dynamics. 
It is known for a long time that the centrocytes
undergoing apoptosis were in cell cycle
a few hours ago (Fliedner, 1967) such that it is likely that they
differentiated from the centroblasts. On the other hand
apoptosis takes place in the light zone
(Hardie et al., 1993) where centroblasts are not present
in high density.
The selected centrocytes are emitted from the GC
with probability $q$ and differentiate either into
antibody producing plasma cells or to memory cells.
The model does not distinguish between plasma and
memory cells. Only their sum is taken into account
and is denoted by {\it output cells} $O(\phi)$.
Nevertheless, the dynamics of both types of
output cells may be different (Choe \& Choi, 1998).
Also, the degree of affinity maturation differs (Smith et al., 1997).
In addition, we do not consider any further proliferation
of output cells in or outside of the GC, 
which may be important for a quantitative
comparison with experimental measurements.

Alternatively the selected
centrocytes are recycled to become centroblasts and
to reenter the proliferation process in the dark zone.
This happens with probability
$1-q$ and contributes to the centroblast distribution, 
corresponding to the last term in \gl{modell}.

We want to emphasize that we do not resolve the
time course of the selection process, which is regarded
to be fast with respect to the centroblast proliferation
time scale. This procedure is justified
by a minimal duration of a typical selection process
of $1-2$ hours (van Eijk \& de Groot, 1999), 
which is about one fourth of the
proliferation time scale. Nevertheless, one should be
aware of possible implications due to the fact that
the selection process does not occur instantaneously.
We effectively enclose the selection time in the parameter
$g$, which in this way
becomes a measure of a complete selection
cycle including the differentiation into centrocytes, a first
selection at the FDCs, a second selection at T-helper
cells and finally the recycling process.
This has two consequences for the model: 
The number of selectable centrocytes for each shape
space state $\phi$ is given at every time by \gl{centrocytes},
i.e.\,the number of centrocytes just being in the selection
process. Furthermore, the details of the multi-step
selection process (Lindhout et al., 1997) are omitted. 
The selection is modeled by a sum over the shape space of
the antigen distribution presented on the FDCs
weighted by the affinity function \gl{affinity}. In other
words the centrocytes $gB(\phi)$ are selected if an 
antigen is close enough in the shape space.
The meaning of {\it close enough} is determined by the
width of the affinity function and by the probability $a_0$
of a positive selection for an optimal centrocyte with
respect to one presented antigen. 

One should be aware
that the probability $a_0$ is not necessarily equal to one, 
as centrocytes are in the state of activated apoptosis (Cohen et al., 1992)
such that their life time is finite and they have to be selected within
this life time to be rescued from apoptosis. This maximum
selection probability is determined by the relation of
the centrocyte life time of about $10\,h$ 
(Liu et al., 1989, 1994)
and the duration of the selection process of $1-2\,h$
(van Eijk \& de Groot, 1999).
A rough estimate using two Gaussian distributions with
reasonable widths for the life time and the selection time
peaked at $10\,h$ and $1.5\,h$ respectively leads to
$a_0\approx 0.95$. Thus, if a GC contains only
centrocytes of optimal complementarity to a presented
antigen, still about $5\%$ of them will undergo apoptosis.

In conclusion, selection can be described by
the convolution term
\be{selection}
\sum_{\phi^*} C(\phi) A(\phi^*) \,a_0 \,a(\phi,\phi^*)
\quad,
\ee
where the number of selectable centrocytes is given by
\gl{centrocytes} and $A(\phi^*)$ represents the distribution of
antigens on the shape space. This selection term
contributes with probability $q$ to the production of 
plasma and memory cells
\be{output}
\frac{dO}{dt}(\phi) \;=\;
q \,\sum_{\phi^*} C(\phi) A(\phi^*) \,a_0 \,a(\phi,\phi^*)
\quad.
\ee
All selected centrocytes not emitted from the GC
are recycled and in this way enhance the centroblast
population giving rise to the last term in \gl{modell}.

\subsection{Initial conditions}
\subsubsection*{Antigen distribution}

Throughout the paper the antigen distribution is assumed
unequal to zero at exactly one point in the
shape space:
\be{antigenini}
A(\phi^*) \;=\; A(\phi_0) \;=\; \delta(\phi^*-\phi_0)
\quad,
\ee
where $\delta(\cdot)$ is one for its argument equal 
to zero and zero otherwise. Consequently, 
the sum in the convolution term
in \gl{modell} is reduced to one contribution.
This implies that we consider only one type of antigen epitope
to be present in the GC.
What seems to be a restriction at first sight
turns out to be an assumption without any loss of generality.
In view of about $10^{11}$ possible states 
in the naive repertoire of humans (Berek \& Milstein, 1988) 
two randomly chosen antigen epitopes
will not be in direct neighborhood of each
others in the shape space.
Both antigen epitopes will only have an influence
on each other during the GC reaction when centroblasts
corresponding to one of the antigen epitopes have a non-negligible
affinity to the other epitope.
In the shape space language this situation corresponds to an 
overlap of two spheres centered at both antigen epitopes.
The radius of these spheres is determined by the maximum
number of mutations which may occur during a GC reaction.
This number will rarely exceed $9$ (K\"uppers et al., 1993;
Wedemayer et al., 1997) such
that in a $4$-dimensional shape space (a more detailed
discussion of the shape space dimension follows in 
Sec.~\ref{ssdim})
one sphere of this size covers about $10^{-8}$th
part of the space. An overlap is unlikely as
an unreasonable large number of $10^8$ antigen epitopes 
is needed to get an important probability for a mutual influence.
Therefore we consider one-antigen epitope GC reactions only. 
A multiple-antigen or a multiple-antigen epitope
GC reaction has to be considered as
sum of one-antigen epitope GCs.

There exists experimental evidence that the amount of presented antigens
is reduced during a GC reaction
(Tew \& Mandel, 1979). Since the amount of antigen
is only halved during $30$ days 
(Tew \& Mandel, 1979; Oprea et al., 2000)
we do not expect an important recoil effect on the
selection probability, which would become time dependent
otherwise. Furthermore, we believe that this antigen diminution
is too weak to be responsible for the vanishing cell
population in GCs after about $3$-$4$ weeks. Anyhow, 
we aim to show that our three-phase model allows for
a vanishing population solely due to the dynamics of the system, 
without the inclusion of any antigen distribution dynamics.

\subsubsection*{B-cell distribution}

The B-cells which have moved to FDCs are in a fast proliferation
phase. In this first phase of GC development the
cell population grows exponentially until it has reached
about $10^4$ cells after three days. One may ask
how many seeder cells are necessary and sufficient
to give rise to this large cell population within such a short
space of time. One finds experimentally that the follicular
response is of oligoclonal character 
(Liu et al., 1991; Jacob et al., 1991; K\"uppers et al., 1993;
Kroese et al., 1987) and the number of
seeder cells is estimated between two and six.
This result is in accordance with the proliferation
rate of $\ln(2)/(6\,h)$ cited above.

The number of seeder cells $\sum_\phi B(\phi,t=0)$ 
determines the initial
B-cell distribution $B(\phi)$ in our model. Corresponding
to the experimental results cited above, we
start with the reasonable number of 
three seeder cells for the follicular response
to an immunization with a specific antigen.
These cells are chosen randomly in the shape space 
within a sphere of radius $R_0$ around the presented
antigen. Due to the assumed affinity neighborhood
this reflects a threshold affinity of antigen and
activated initial B-cells. This was 
already found experimentally (Agarwal et al., 1998):
The activation of a B-cell by
an antigen is necessarily connected with a minimum
affinity of the corresponding epitopes and
antibodies. The radius of the sphere can be
determined by the maximum number of mutations
occurring during the process of affinity maturation
in GCs: 
\be{r0}
R_0=N_{\rm max}
\quad,
\ee
which is found to be $8$ or $9$ 
(K\"uppers et al., 1993; Wedemayer et al., 1997).
In this picture, B-cells with an affinity above threshold
are activated by a specific antigen, giving
rise to a fast immune response with antibodies
of low but non-vanishing affinity. The
germinal center reaction has to be understood
as an optimization process for these initially activated
B-cells leading to a second slower primary immune 
response of high specificity.

\subsection{Width of affinity function}

Due to affinity neighborhood in
the shape space the affinity function may be chosen
according to \gl{affinity}. 
The width $\Gamma$ has to be consistent with the
affinity enhancement that is achieved during a GC 
optimization process.
Let the factor of affinity enhancement be $10^\xi$
and the number of corresponding somatic hypermutations
be $N$. Then from \gl{affinity} we get
\be{width}
\Gamma \;=\; \frac{N}{\left(\xi \ln(10)\right)^{\frac{1}{\eta}}}
\quad.
\ee
Typical numbers of somatic hypermutations are $N=6$
(K\"uppers et al., 1993) corresponding to an affinity enhancement
of $100$ ($\xi=2$) (Janeway \& Travers, 1997), while $N=8$ 
is considered to be a large number of
mutations (K\"uppers et al., 1993)
corresponding to a high factor of affinity enhancement 
of $2000$ ($\xi=3.3$) (Janeway \& Travers, 1997). A huge
affinity enhancement by a factor of $30000$ ($\xi=4.5$)
is reached with $N=9$ mutations (Wedemayer et al., 1997). 
As can be
seen in \abb{Gamma}, these values for the affinity
enhancement with growing number of somatic
hypermutations are consistent with a Gaussian affinity
function ($\eta=2$) with width of $\Gamma\approx 2.8$.
\begin{figure}[ht]
\vspace{-1cm}
\center{\hspace*{0mm} \epsfxsize=10cm \epsfysize10cm
        \epsfbox{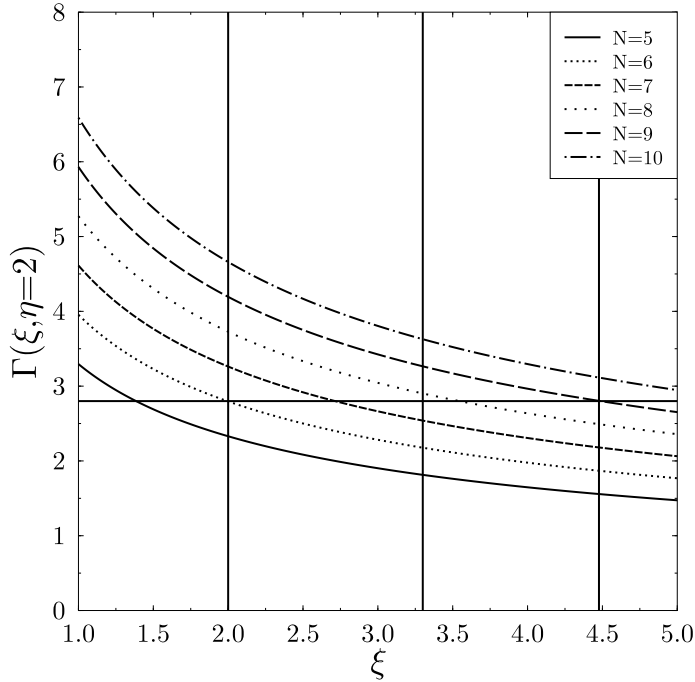}}
\caption[]{\sf The width of the Gaussian affinity function 
as a function of the factor of affinity enhancement 
for different numbers of somatic hypermutations.
Vertical lines denote the positions of the used
experimental values of affinity enhancement. One
can observe that consistently for all three values the resulting
width becomes approximately $2.8$.}
\label{Gamma}
\end{figure}
It should be mentioned that a consistent value
for the width of the affinity function can not be
obtained for other integer values of $\eta$.
But one should be aware that the used
affinity enhancements are very rough estimates and
thus our result should be understood as a first
guess -- even if the compatibility with the experimental
data is convincing.

\subsection{Shape space dimension}
\label{ssdim}

The shape space dimension $D$ enters the model through
the mutation term in \gl{modell} and has to be chosen
appropriately to our model. 
We already mentioned that somatic hypermutation is likely to
occur randomly (Weigert et al., 1970; Radmacher et al., 1998), 
i.e.\,that the direction of mutation in the shape space
is not governed by the position of the antigen.
The term describing mutation in \gl{modell} is in
analogy to a diffusion term in a continuous space
as used in (Perelson \& Wiegel, 1999). 
Let us assume for a moment that the
development of the GC reaction is governed 
by a diffusion process
and let the number of maximum point mutations
be $N_{\rm max}$. Then the diffusion process
operates in a sphere around a seeder cell
of radius $N_{\rm max}$
in the shape space. In order to match the position
of the antigen with at least one centroblast, all positions
in the sphere have to be reached by diffusion.
This means that the total number of
centroblasts $N_c$ present in the GC $3$ days after immunization, 
i.e.\,when selection is triggered, should be of the
order of magnitude of the number of points in the sphere.
The number of points in the sphere grows exponentially
with the dimension of the shape space.
Thus, for $N_{\rm max}=8$ and $N_c=12000$ it follows that
the shape space dimension should be $4$. From the point
of view of a pure diffusion process this is an upper bound for
the shape space dimension.

However, the above argument does not include
the proliferation rate, the selection rate, and the
recycling probability. 
The effective number of centroblasts 
available to reach the antigen may become larger than the
$12000$ cells assumed above through proliferation and
recycling. As we want to describe a decreasing population
for large times without further assumptions,
the effective rate of production of additional centroblasts
during the GC reaction cannot
exceed some small value, since the centroblast population would
explode otherwise for large times. So it seems unlikely that
the shape space dimension appropriate for our model is
substantially larger than $4$. We will start with
$D=4$ and check our results for the
dimensions $5$ and $6$.
We want to remind that we
are dealing with a phenotype space in this model and that
a genotype space probably requires a substantially
higher dimension.

\subsection{Long term behavior}
\label{gvalue}

The parameter $g$ was introduced in the model as the
rate of differentiation from centroblasts in the dark zone
into antibody presenting centrocytes ready to
undergo a selection process. 
As we do not consider
the centrocyte dynamics during the multi-step 
selection process (Lindhout et al., 1997),
this parameter effectively describes the total duration
of the selection process including the differentiation
mentioned above, the selection at the FDCs and
with T-helper cells, and the recycling to centroblasts.
Only those cells that have successfully finished this program
are contributing to the centroblast population described
in \gl{modell}.

Thus, being an effective measure of the speed of the
selection process, we are able to deduce an upper
bound for $g$. Solely the inhibition of apoptosis
during the selection process at FDCs and
with T-helper cells takes at least $1$-$2$ hours 
(Lindhout et al., 1995; van Eijk \& de Groot, 1999).
This gives us an upper bound for the rate $g$ of
\be{gupper}
\frac{g}{\ln(2)} \;<\; \frac{1}{2\,h}
\quad.
\ee
Nevertheless, this upper bound is not sufficient to
fix $g$. As $g$ governs the whole reaction speed it will
play a crucial role for the final state $21$ days after 
immunization. As we know that at the end of the GC
reaction only a few cells remain in the environment of 
the FDCs (Liu et al., 1991; Kelsoe, 1996),
the parameter $g$ will be tuned in order to get this result.

Mathematically, this requirement has the form of a
long term condition. Let us look for an equilibrium
condition at the point $\phi_0$
representing the antibody type of optimal complementarity to
a presented antigen:
\be{equilibrium}
\frac{dB}{dt}\left(\phi=\phi_0,t\right)\;=\;0
\quad.
\ee
Then, the selection term in \gl{modell}
becomes simply $(1-q)ga_0 B(\phi_0,t)$ and
for a non-vanishing centroblast distribution 
at $\phi_0$ we get
\be{equilcondition}
\eps\;\equiv\;
p-2pm-g+\frac{pm\beta}{D}+(1-q)ga_0\;=\;0
\quad,
\ee
where we have introduced $\eps$ as a measure for the
vicinity to the equilibrium condition and
\be{betadef}
\beta(\phi_0,t) \;=\; 
\frac{\sum_{||\Delta\phi||=1} B(\phi_0+\Delta\phi,t)}
       {B(\phi_0,t)}
\quad.
\ee
$\beta(t)$ is a measure for the sharpness of the centroblast
distribution in the shape space around the antigen position
$\phi_0$. 
$\beta$ becomes constant when the
selection process starting at day $3$ after immunization
has reached a phase in which the
centroblasts at the position of the antigen dominate.
This is already the case $8$ days after immunization
according to experiments (Jacob et al., 1993) and
is confirmed in our model (see \abb{beta46}).

Thus, at large times $\beta$ can already be considered as
constant in the equilibrium condition \gl{equilcondition} 
and the same applies to $\eps$. 
In this stage of GC development
the centroblast population at $\phi_0$ has a pure 
exponential form
\be{Bexp}
B(\phi_0,t) \;=\; e^{\eps t}
\quad.
\ee
The sign of $\eps$ determines whether the centroblasts
population explodes ($\eps>0$) or dies out
($\eps<0$). 
Because of the finite life times of GCs, $\eps$
should adopt values slightly smaller than zero.

\subsection{Recycling probability}

The recycling probability $1-q$ determines the fate of the
positively selected centrocytes in the light zone. Apoptosis
was inhibited for these cells and they may either become
plasma or memory cells or return to the dark zone to reenter
the fast proliferation phase. This recycling hypothesis 
(Kepler \& Perelson, 1993)
has been intensively discussed and the main position is
that random somatic hypermutation
-- and it is likely that somatic hypermutation 
occurs randomly 
(Weigert et al., 1970; Radmacher et al., 1998), 
i.e.\,that the direction
of mutation in the shape space does not depend on the
position of the antigen
-- does not lead to a sustained optimization of affinity
in a one-pass GC reaction (Oprea et al., 2000).
To reach a specific
position in the shape space a multi-step optimization and 
verification process is required to avoid an arbitrary aimless
walk through this high dimensional space.

To propose a new perspective onto this question we consider the most
convincing experimental evidence for the existence of such a
recycling process (Han et al., 1995b). A GC reaction was
initiated in mice with one antigen. $9$ days after immunization, 
i.e.\,in a stage
of the reaction in which the {\it good} cells already dominate
(Jacob et al., 1993), a second
different, but related antigen was injected. 
This change of the reaction conditions 
in the course of the GC reaction leads to some very interesting
observations: The rate of apoptosis of centrocytes is enhanced by a factor
of about $\nu\approx 5$ compared to the non-disturbed GC reaction.
This can be explained by the fact that the antigen distribution 
in the shape space
was changed such that most of the cells present
after $9$ days may fit to the first antigen but not to the
second one. The probability of a positive selection is diminished and
the inhibition of apoptosis does not occur to the same extent
as before. On the other hand the total cell population in
the GC vanishes with a $\omega\approx 8$-fold higher 
speed compared to the unperturbed GC reaction. 
This is very difficult to explain without
recycling: As we know that the centrocytes (only very small
numbers of centroblasts) undergo apoptosis
if not rescued in time, the change of the apoptosis rate
is of relevance only for the centrocyte population. The faster
reduction of the whole GC cell population can not be explained
by the enhanced apoptosis rate, if there exists no recoil effect
of the centrocyte apoptosis to the centroblast population.
This is an indirect confirmation of the recycling hypothesis.

We want to consider this argument in
a quantitative way by translating this experiment
into the language of our model. We know from our
consideration in Section \ref{gvalue} that
at day $9$ after immunization the optimization process
is already completed and that the centroblast population
at the antigen position $\phi=\phi_0$
in the shape space behaves according to \gl{Bexp}.
But due to the modification of the antigen distribution
at day $9$ after immunization the GC reaction returns 
back to its dynamical
selection phase. Thus, until day $9$ the centroblast
population $B_1(\phi,t)$ evolves with respect to
the antigen distribution $A_1(\phi^*)$ 
as in \gl{antigenini}
and after day $9$ two GC developments have to
be compared: One continues its usual reaction
as before. The second $B_2(\phi,t)$ 
starts with the distribution
$B_1(\phi,t=9\,d)$ and evolves with respect to
the new antigen distribution
\be{anti2}
A_2(\phi^*)\;=\;
\rho_1 \delta(\phi^*-\phi_0)
+\rho_2 \delta(\phi^*-(\phi_0+\Delta\phi^*))
\quad,
\ee
where $\Delta\phi^*$ denotes the shift of the additional
second antigen with respect to the first one. $\rho_{1,2}$
take into consideration the (possibly) different concentrations 
of both antigens.

Now we are able to calculate the apoptosis rate in both
scenarios for the cells of some fixed but arbitrary type $\phi$. 
The number of centrocytes available for
selection at every moment is $gB(\phi)$. The apoptosis
rate is given by this number reduced by the selected
cells
\be{apoprate}
V_i(\phi) \;=\; 
gB(\phi)
\left(1-a_0\sum_{\phi^*} A_i(\phi^*) a(\phi-\phi^*)\right)
\quad,
\ee
where $i$ denotes the two scenarios. For the original
GC reaction this becomes
\be{apopeins}
V_1(\phi)
\;=\;
gB(\phi) \left(1-a_0 a(\phi-\phi_0)\right)
\quad,
\ee
while in the GC with the new antigen distribution
we find
\be{apopzwei}
V_2(\phi)
\;=\;
gB(\phi)
\left(1-a_0 a(\phi-\phi_0) \rho_1
          \left(1+ \alpha(\phi)\right)
\right)
\quad,
\ee
where we have shifted $\phi^*$ in the second term
of \gl{anti2} and introduced the ratio
\be{alpha}
\alpha(\phi) \;=\; 
\frac{\rho_2 a(\phi-\phi_0-\Delta\phi^*)}
{\rho_1 a(\phi-\phi_0)}
\ee
valid for each point $\phi$ separately.
The factor $\nu$ of the apoptosis enhancement found
in the experiment above is then given by the ratio of
both apoptosis rates
\be{nu}
\nu \;=\;
\frac{V_2}{V_1} \;=\;
\frac{1-a_0 a(\phi-\phi_0) \rho_1
            \left(1+ \alpha(\phi)\right)}
{1-a_0 a(\phi-\phi_0)}
\quad.
\ee

On the other hand, this enhancement of apoptosis leads
to a faster population reduction by a factor $\omega$,
which in our model is given by the ratio of the changes
of the cell distributions over the shape space
at the time of the presentation of the new antigen.
According to \gl{modell} and using the new
antigen distribution $A_2$ this ratio is given by
\bea{omega}
\omega 
&=&
\frac{dB_2}{dt}
\left[\frac{dB_1}{dt}\right]^{-1}(\phi,t=9\,d)
\nn
&=&
\frac{p-2pm-g+\frac{pm}{D}\,\beta(\phi)
       +(1-q)ga_0 a(\phi-\phi_0)\,
          \rho_1 \left(1+\alpha(\phi)\right)}
{p-2pm-g+\frac{pm}{D}\,\beta(\phi)
       +(1-q)ga_0 a(\phi-\phi_0)}
\quad,
\eea
where the cell distributions $B(\phi,t=9\,d)$ cancel as they
are equal in both scenarios for $t=9\,d$.

Taking the results for $\nu$ (\gl{nu}) and for $\omega$ (\gl{omega})
at the shape space point of
the primary antigen $\phi=\phi_0$
the affinity function becomes equal to one. $\beta$
is already a constant for $t=9\,d$ such that we may
eliminate $\rho_1(1+\alpha(\phi_0))$ from both equations
to get
\be{nutoomega}
\frac{\nu-1}{\omega-1}
\;=\;
\frac{p-2pm-g+\frac{pm}{D}\,\beta(\phi_0)+(1-q)ga_0}
      {(1-q)g(a_0-1)}
\ee
or for the recycling probability
\be{recycle}
1-q \;=\;
-\frac{p-2pm-g+\frac{pm}{D}\,\beta(\phi_0)}
{g\left[a_0+(1-a_0)\frac{\nu-1}{\omega-1}\right]}
\quad.
\ee
This relation allows us to determine the parameter $q$ and in
this way to give a quantitative prediction for the recycling 
probability for selected centrocytes to reenter the dark zone and to
continue proliferation. 
The recycling probability has to fulfill \gl{recycle}
in order to be consistent with the experiment 
of Han et al., (1995b).
We want to emphasize, that the condition \gl{recycle} is independent
of the concentrations of the two antigen types used in the
experiment.

\subsection{Output dynamics}
\label{outputdyn}

The production of plasma and memory cells optimized with
respect to the presented antigen in
the course of the GC reaction is governed by
\gl{output}. According to our discussion of the different phases
of the GC reaction in \kap{dynamics}, the starting time
of {\it output cell} production is not necessarily
correlated with the starting time of the mutation
and selection process about $3$ days after immunization. 
There may occur
a time delay of the output production in relation to the
selection process (see Fig.\,\ref{phases}). 
This possibility is encountered by 
comparing the dynamics of 
output cell production with experiment 
(Han et al., 1995a).
Here, the number of output cells of high affinity to the presented
antigen $\phi_0$ is compared at day $6$ and
day $12$ after immunization and their relation is found
to be approximately 
\be{out126}
v_O
\equiv
\frac{\int_{0d}^{12d}dt\,O(\phi_0)}
       {\int_{0d}^{6d}dt\,O(\phi_0)}
\approx
6
\quad.
\ee
As output cells are already present at day $6$ after immunization,
the starting point of output cell production has to fulfill
$0\,h<\Delta t_2<72\,h$ 
($\Delta t_2$ denotes the time window
between the start of mutation and selection and the start
of output cell production, see Fig.~\ref{phases}).
It is intuitively clear that for large $\Delta t_2$ the speed
of output cell production $v_O$ will be enhanced, for 
the cell population of the GC will already be peaked at the 
position of the antigen in the shape space. Thus, we 
are able to adjust $\Delta t_2$ to the correct 
output cell dynamics.


\section{Results and Predictions}

\begin{table}[ht]
\begin{center}
\begin{tabular}{|l|r@{$\;=\;$}l|} \hline
Parameter & quantity & value \\ \hline
Proliferation rate & $p/\ln(2)$ & $1/(6\,h)$ \\
Somatic hypermutation probability & $m$ & $0.5$\\
Rate of differentiation of centroblasts to centrocytes & $g/\ln(2)$ & 0.352\\
Dimension of shape space & $D$ & 4\\
Selection probability for optimal centrocytes & $a_0$ & $0.95$\\
Output probability of selected centrocytes & $q$ & 0.2\\
Number of seeder cells at FDCs & 
$\sum_\phi B(\phi,t=0)$ & $3$\\
Radius of B-cell activation around antigen & $R_0$ & $8-10$\\
Time duration of phase I of GC reaction &
$\Delta t_1$ & $72\,h$\\
Time duration of phase II of GC reaction &
$\Delta t_2$ & 48\,h\\
Time duration of the whole GC reaction &
$\sum_{i=1}^3 \Delta t_i$ & $504\,h$\\
\hline
\end{tabular}
\caption[]{\sf Summary of all parameters of the model.
They were determined by experimental data. 
For explanations and references see the last section. 
The parameters $g$, $q$, and $\Delta t_2$ are determined 
in the course of the solution of \gl{modell} with respect to experimental 
data discussed before.}
\label{parameter}
\end{center}
\end{table}

\begin{figure}[ht]
\center{\hspace*{0mm} \epsfxsize=10cm \epsfysize10cm
            \epsfbox{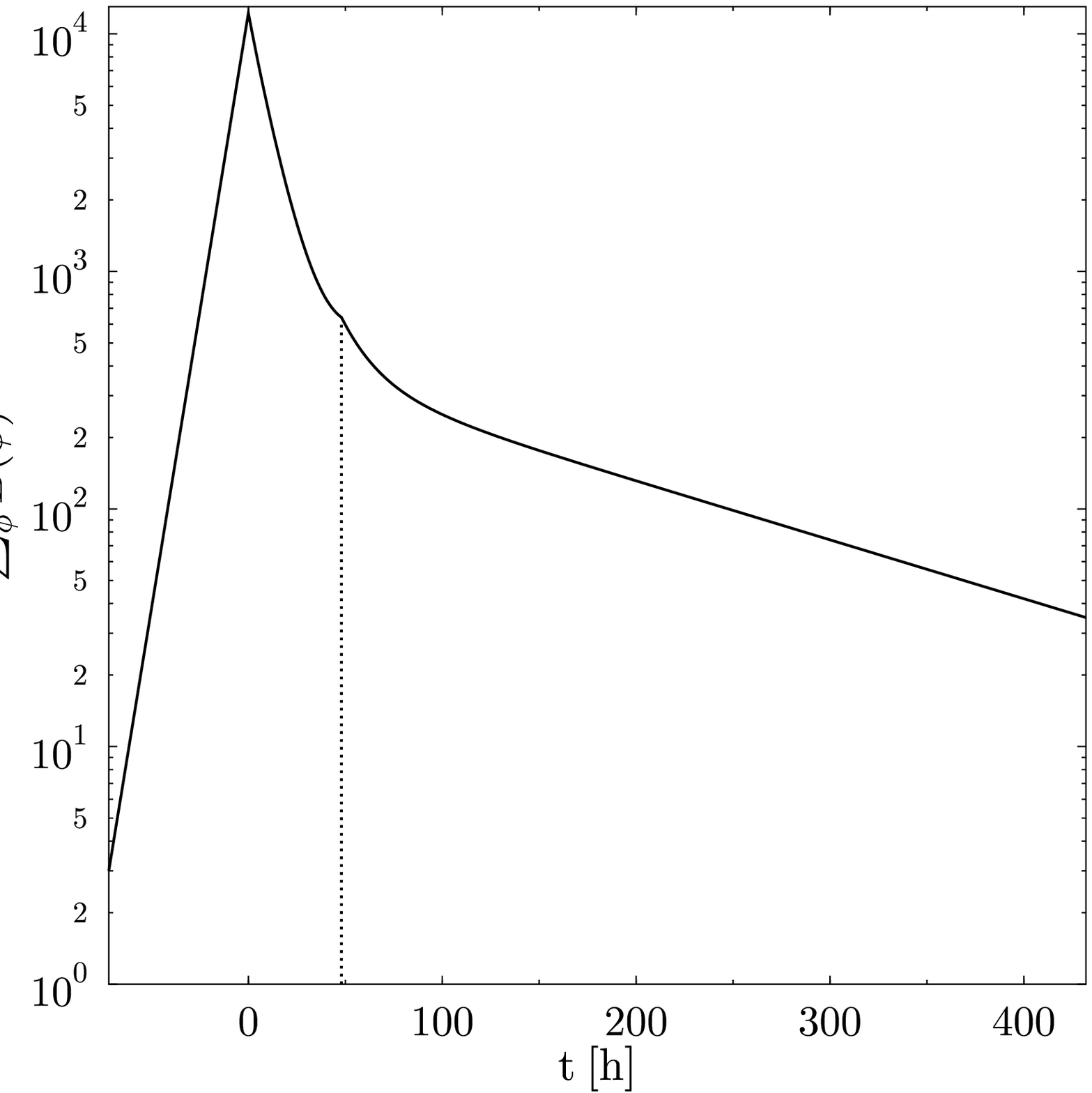}}
\caption[]{\sf Time course of the centroblast population integrated 
over the whole sphere in the shape space around the antigen
$\phi_0$. In the first phase the initial B-cells proliferate. The
large population reached at $t=0\,h$ is then reduced by the
selection process. At
$t=48\,h$
the production of plasma and
memory cells starts, leading to an exponential decrease in
the overall centroblast population.}
\label{bsum46}
\end{figure}
Starting from a randomly chosen antigen epitope and --
following the oligoclonal character of the B-cell population 
of a GC -- three randomly chosen activated B-cells in a
sphere of radius $R_0$ according to \gl{r0}, we let the
GC reaction develop as described by \gl{pmodell} and
\gl{modell}.
The different phases (see \kap{dynamics}) of the GC reaction
are respected. The parameters are determined from experimental
data as summarized in \tab{parameter}. The parameters
$g$, $1-q$ and $\Delta t_2$, describing the differentiation rate
of centroblasts to centrocytes, the recycling probability for
the selected centrocytes and the time delay for the production
of output cells, respectively, are iteratively adjusted 
in order to fit with the experiments
described in the previous section. The time variable 
is running from $t_0=-72\,h$ to accentuate the first phase of 
pure proliferation as a {\it preGC-phase}. The mutation and
selection process is started at $t=0\,h$ and the production
of plasma and memory cells begins at $t=\Delta t_2$. The
differential equation is solved numerically with a modified Euler
method in a subspace, in which we assume Dirichlet boundary
conditions with $B=0$.

The best results are obtained by an iteration procedure for
\be{bestpar}
\frac{g}{\ln(2)}=\frac{0.355}{h} 
\quad , \quad
1-q=0.8
\quad ,\quad
\Delta t_2=48\,h
\quad.
\ee
At $t=0\,h$ we find a large oligoclonal population of 
$12288$ centroblasts as a result of the proliferation phase
started with $3$ initial cells 
(see \abb{bsum46}). These cells are of different type compared to
the cells of optimal affinity to the presented antigen and in a 
typical distance from them (varying between
$3$ and $8$ point mutations to reach the optimal cell type).

\paragraph{Optimization phase}
The selection process at the FDCs and with the T-helper cells
starts and reduces the centroblast population with low affinity
to the presented antigen. At the same time
somatic hypermutation
leads to a spreading of the centroblast distribution
over the shape space, resulting in a small but non-vanishing amount
of centroblasts at the shape-space position of the antigen
(see \abb{boe46}).
\begin{figure}[ht]
\center{\hspace*{0mm} \epsfxsize=10cm \epsfysize10cm
            \epsfbox{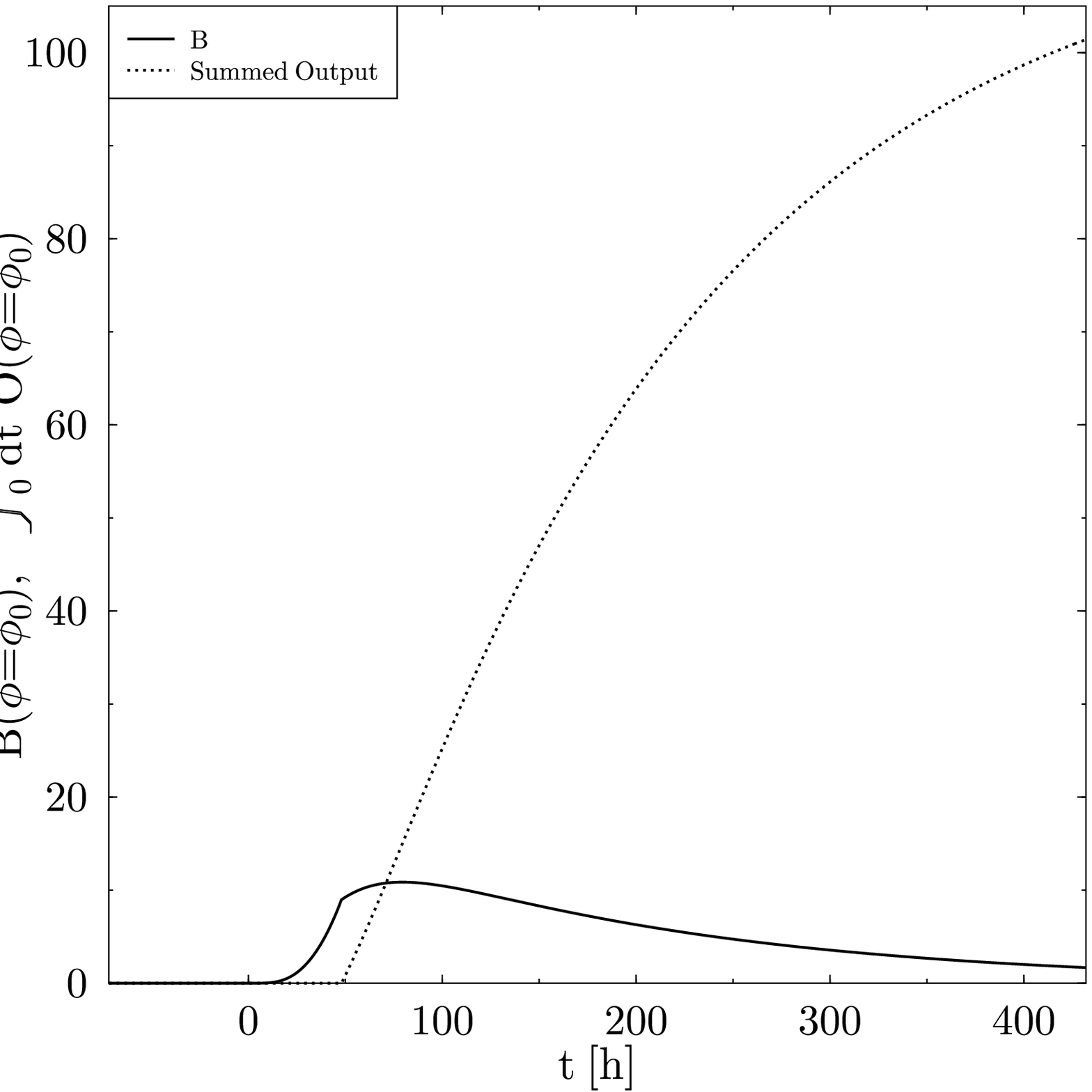}}
\caption[]{\sf The centroblast population and the 
integrated plasma/memory cell production at the position
of the antigen in the shape space. The centroblast population
shows a growth due to proliferation and
somatic hypermutation starting at $t=0\,h$. 
The population growth is slowed down by
the production of output cells starting at $t=48\,h$. Now
the number of produced plasma and memory cells is increased
steadily until the end of the GC reaction. Then, only
$2$ cells (of optimal type) remain in the environment of the 
FDCs.}
\label{boe46}
\end{figure}
The large majority of B-cells does not survive the 
selection process (compare the summed cell population 
decrease in \abb{bsum46}), dies through apoptosis and
is rapidly phagocytosed by macrophages present in the GC.

\paragraph{Secondary proliferation}
As all positively selected centrocytes reenter the proliferation
process in the optimization phase of the GC reaction, the number of 
centroblasts at the antigen position grows. Note
that $\eps>0$ follows for the equilibrium condition \gl{equilcondition}
because of $q=0$, according to \gl{phase2}. As a consequence,
the overall centroblast population would restart to grow
if this growth was not inhibited by the production of plasma 
and memory cells after $48\,h$. Especially, 
this {\it secondary proliferation} process is found for the centroblasts
encoding antibodies of high affinity to the antigen
(see \abb{boe46}). The centroblasts at the antigen position
are practically all recycled (beside the reduction due to
$a_0<1$) such that no reduction process exists any more.
This gives rise to a new perspective on the GC reaction:
We need just as many initially proliferated cells, that a
spreading of the distribution through somatic hypermutation
leads to at least one or two cells matching the antigen position in the
shape space. Then a considerable amount of cells of this
optimized type is reached within the time delay $\Delta t_2$ of the production
of output cells, opening the possibility of a full proliferation
process for these cells.

\paragraph{Recycling probability}
When the output production is turned on at $t=48\,h$
the dynamics of the germinal center is characterized by
$\eps<0$ in \gl{equilcondition}, such that an exponential decrease
of the whole centroblast population is initiated -- including
the population of the optimized cells 
(see \abb{bsum46} and \abb{boe46}). 
Nevertheless, this decrease is slow enough
to allow the production of a considerable amount of plasma
and memory cells (see \abb{boe46} and \abb{showo46}).
\begin{figure}[ht]
\center{\hspace*{0mm} \epsfxsize=10cm \epsfysize10cm
            \epsfbox{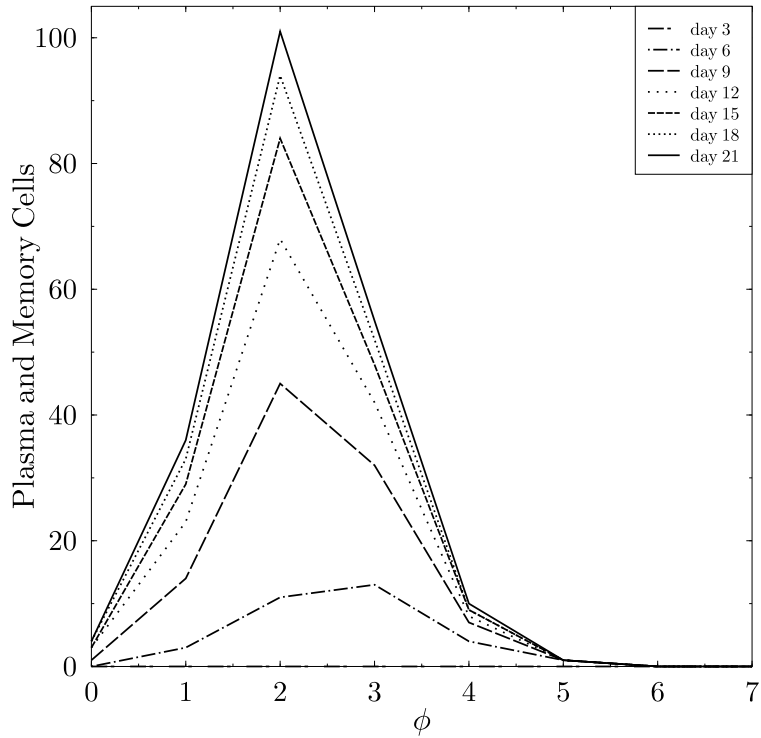}}
\caption[]{\sf A cut through the sphere in the shape space around the 
presented antigen at position $2$. The seeder cells are at a distance of
at least $5$ mutation steps.
The number of produced plasma and memory
cells grows considerably during the GC reaction. The
rapidity of growth diminishes in course of time because of the decreasing
number of optimized centroblasts at the antigen position.}
\label{showo46}
\end{figure}
It is an interesting observation that a rather small number of 
centroblasts of the optimal type with respect to the antigen
is sufficient to produce this large number of plasma and 
memory cells -- even with a small output probability of
$q=0.2$. On the contrary, a larger output probability may
decrease the integrated number of plasma and memory cells,
as less cells reenter the proliferation process. In conclusion,
we claim that a large recycling probability of $80\%$ 
of the positively selected centrocytes is
not only necessary to achieve an optimization by 
multi-step mutations but also to achieve a high
number of resulting plasma and memory cells.

\paragraph{The end of the GC reaction}
Finally, at day $21$ after immunization, only a few cells
with maximum affinity to the antigen remain
in the environment of the FDCs, as required from experiment 
(Liu et al., 1991; Kelsoe, 1996). So the dynamics of our
model allow for a vanishing of the germinal center population
without any reduction of the antigen concentration or other
additional requirements. The length of the
germinal center reaction is determined by the interplay of 
the centroblast (into centrocyte) differentiation rate $g$ and the
delay of output production $\Delta t_2$, which is determined
by \gl{out126}, so that the GC
life time is basically controlled by $g$. 
So from the perspective of our model a diminishing cell
population may result solely
from the interplay of proliferation, differentiation of
centroblasts to centrocytes, and output production.

\paragraph{Dependence of initially activated seeder cells}
The total number of remaining cells depends on the 
distance of the initial B-cells from the antigen $\phi_0$.
For numbers of mutation steps (necessary to reach the antigen)
between $N=3$ and $N=8$, the number of remaining
B-blasts varies from $43$ to $0$. For a typical number
of mutations of $N=5$ about $10$ B-blasts remain in the
environment of the FDCs at day $21$ after immunization.
In other words, the GC life time depends on the initial
distribution of the seeder cells with respect to the presented
antigen. It would be interesting to check this statement
experimentally by observing GCs with a variable number of
mutations occurring during the optimization process.
One could imagine for real GCs a self-regulating process 
which depends on the already positively selected centrocytes and thus
prolongs the GC reactions with large distances between
seeder cells and antigen, and shortens correspondingly
the ones with small distances.
However, there is experimental evidence that the intensity of the
GC reaction depends on the quality of the initially activated
B-cells (Agarwal et al., 1998). Also in quasi-monoclonal mice
(Cascalho et al., 1996) it was found that the volume of
GCs 4 days after immunization depends on the average affinity
of the reservoir B-cells to the antigen (de Vinuesa et al., 2000,
Fig.~6). A quantitative comparison of these data with our model results
would be interesting.

\subsection{Result stability}

The parameter set in \gl{bestpar} is determined such 
that the experimental conditions \gl{recycle}
and \gl{out126} are fulfilled for typical initial conditions. 
Therefore, the prediction
of the recycling probability to be $80\%$ is a statement
which -- in the framework of our model --
is justified to the same degree as the experiments
are exact. 
Nevertheless,
the sensitivity of the predicted high recycling probability
to such modifications is weak.

It should be
mentioned that in addition there is a certain freedom of 
choice concerning the number of remaining cells at day
$21$ after immunization. For example, the parameter set
with a still higher recycling probability
\be{par2}
\frac{g}{\ln(2)}=\frac{0.52}{h} 
\quad , \quad
1-q=0.9
\quad ,\quad
\Delta t_2=55\,h
\quad.
\ee
leads to a result of comparable quality as under the conditions
of \gl{bestpar}
-- it is clear that a higher recycling probability slows
down the production of output cells such that simultaneously
the time delay of the production start has to be larger --
with the difference that the number of remaining cells
is reduced with respect to 
the previous result such that no cells remain at the end
of typical GC reactions. For this reason and because
$g$ is on its upper bound \gl{gupper},
this parameter set was not considered as reasonable.

If one requires a smaller recycling probability the
parameter set
\be{par3}
\frac{g}{\ln(2)}=\frac{0.286}{h} 
\quad , \quad
1-q=0.7
\quad ,\quad
\Delta t_2=42\,h
\quad.
\ee
leads to correct values for $v_0$ and 
\gl{recycle} is also fulfilled.
The reduced recycling probability accelerates the
output cell production such that it has to be decelerated
by a reduction of the time delay $\Delta t_2$. Even if
the differentiation rate stays in a reasonable range,
the number of remaining cells in the GC at day $21$ 
after immunization is about ten-fold with respect to
\gl{bestpar}. More generally one may consider the number
of remaining cells for typical GC reactions with an average
number of mutations necessary to reach the antigen.
\begin{center}
\begin{tabular}{|l|c|c|c|c|c|} \hline
Recycling probability $1-q$ & $0.5$& $0.6$& $0.7$& $0.8$& $0.9$\\
\hline
$\sum_\phi B(\phi,t=21\,d)$ & 1476& 515& 134& 10& 0\\
\hline
\end{tabular}
\end{center}
The mentioned experimental data are all reproduced for each
value of $1-q$, solely the number of remaining cells is left open.
To ensure that this number is in accordance 
with the experimental statement that a {\it a few} proliferating
B-blasts remain at the FDCs at this stage of GC development,
we are led to a recycling probability of $80\%$. We want
to emphasize that this result is determined very clearly, as
for the other recycling probabilities the remaining number of
cells in the GC final state differs by at least an order of magnitude.

This variation of parameters shows the range of
stability of our prediction:\\
First, the window of
possible time delays $\Delta t_2$, which are in
accordance with \gl{out126} is very narrow.
Already for $\Delta t_2>55\,h$ the condition
\gl{gupper} will be violated. On the other hand
for $\Delta t_2<38\,h$, despite the immense 
number of remaining cells in the final GC state,
the condition \gl{out126} can not be fulfilled
anymore. In consequence, it is not possible
that the production of plasma and memory cells
already starts with the establishment of light
and dark zone, i.e.\,that it starts with the existence of
a fully developed GC. The emphasized
experimental bounds require necessarily a time
delay for the production of output cells of
at least $42\,h$. This means that the production
process should be initiated by a separate signal
and is not present in a working GC automatically.\\
Secondly, the window of possible recycling
probabilities is strongly bound by the
experiments (Han et al., 1995a, 1995b) and by the
experimentally observed number of remaining
cells at day $21$ after immunization. This leads
to the conclusion that the recycling probability
should be at least $70\%$ and not larger
than $85\%$. 
Even ignoring the required number of cells in the
final state, \gl{out126} does not allow a 
recycling probability smaller than $60\%$.
We would like to emphasize that this
result differs extremely  from the value for the recycling
probability of $0.15$ assumed in (Kesmir \& de Boer, 1999). 
Nevertheless,
we do not regard this discrepancy as a contradiction
because in the model of Kesmir \& de Boer not all parameters
were fixed by experimental data.

\paragraph{Shape space dimension}
The analysis presented above
was performed with a shape space dimension of $4$,
(compare \kap{ssdim}). It is an important remark that our
results remain essentially unaffected for $D=5$ or $6$.
Using for instance $D=5$, only the parameter
$g$, controlling the selection speed,
has to be reduced slightly by $7\%$ in order to stay in
accordance with the experimental bounds. The total
number of remaining cells at day $21$ after immunization
does not change but its distribution is spread out on the
shape space.
For $D=6$, the
parameter $g$ is again reduced by $6\%$, while all
other parameters remain unchanged to stay in accordance
with the experimental data.
Again, the cell distribution at day $21$ after immunization
becomes less centered at the antigen
position. We conclude that, for the values tested,
the dependence on the
shape space dimension is weak enough such that
our results are not affected by it. Even, the slower
selection speed for higher dimensions of less than
$10\%$ per dimension remains in the range of
accuracy of our results. 

Note, that the dependence on the shape space dimension
is restricted to an intermediate stage (phase II) of the GC reaction.
In the first phase of pure proliferation $D$ does not
enter the dynamics at all.
In the late phase of selection,  
when the distribution $B(\phi)$ becomes approximately
spherically symmetric around the point representing the
antigen, $D$ cancels exactly with the factor $1/D$ in
\gl{modell} as there are $D$ equal terms contributing to
the sum such that $D$ factorizes.
A dependence of our results on the shape space
dimension may only occur during the early phase of selection,
in which the distribution $B(\phi)$ is clearly not symmetric
around the position of the antigen.
Since the distribution of centroblasts
develops into a spherically symmetric distribution around $\phi_0$
already before day $8$ after immunization (see \abb{beta46}),
we deduce that a dependence on the shape space dimension
may occur between day $3$ and day $8$ of the GC reaction only.

\paragraph{Mutations with large changes of affinity}
We have assumed so far that mutations are represented in the
shape space by next-neighbor jumps.
To check the robustness of the result for the recycling probability
against the possibility of mutations that lead to large distance
jumps in the shape space, we consider a worst case argument.
Let us suppose that $10\,\%$ of the mutations lead to a large jump
in the shape space. This is translated into the model by reducing
the number of next-neighbor-mutations by a factor of $0.9$. The
result is a lower efficiency of the mutation process
as the probability for a random mutation to improve affinity
with respect to the antigen is very small. Therefore, the final
number of cells is reduced by a factor of 10, and the
centroblast to centrocyte differentiation rate has to be adjusted
(from $0.244$ to $0.255$) in order to fulfill condition \gl{recycle}.
A similar result as before is obtained by an adjustment
of the recycling probability from
$0.8$ to $0.78$. As this is a worst case argument, we consider
our result as robust against large jump mutations.

\subsection{Optimization speed}

\begin{figure}[ht]
\center{\hspace*{0mm} \epsfxsize=10cm \epsfysize10cm
            \epsfbox{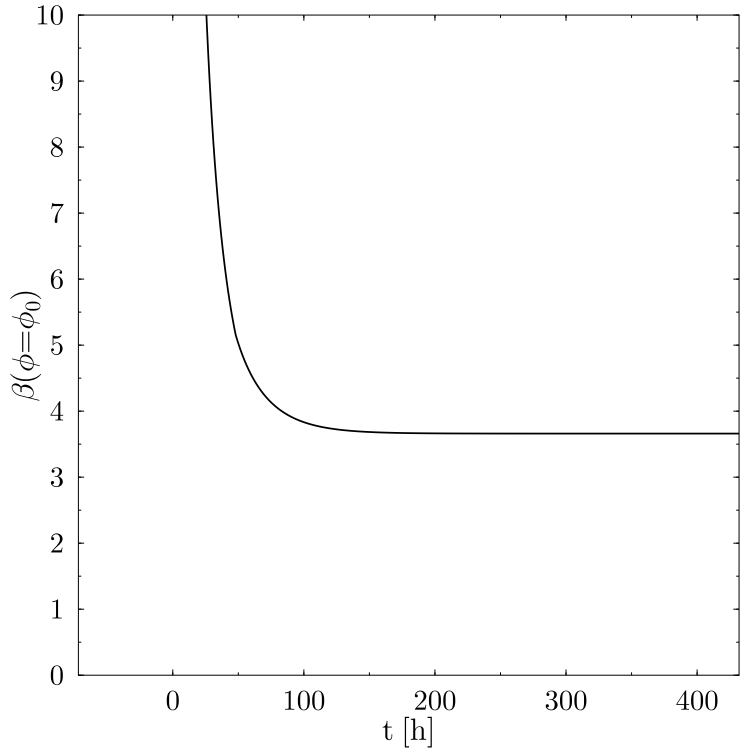}}
\caption[]{\sf The time course of $\beta(\phi_0,t)$ (see \gl{betadef}) 
at the shape space position of the antigen for the parameter set
\gl{bestpar}.
$\beta$ becomes constant shortly after $t=96\,h$, 
i.e.\,at day $7$ after immunization.}
\label{beta46}
\end{figure}
From experiment it is expected that the cells
with high affinity to the antigen dominate already
at day $8$ after immunization (Jacob et al., 1993). 
This is verified in the model by checking if the asymptotic regime
of the GC reaction is reached at day $8$ in the sense that $\beta$
(defined in \gl{betadef}) becomes constant.
As can be seen in \abb{beta46}, $\beta$ becomes
constant in the course of day $7$ after immunization
for the parameter set \gl{bestpar}. This behavior of $\beta$
is not altered in a great range of parameter variation. 
Besides the reproduction of the experimental evidence
for the optimization process to be accomplished after
$8$ days, this result is an a-posteriori
check for the derivation of the recycling
probability, which is based on a constant value
of $\beta(\phi_0)$ at day $9$ after immunization
(see \gl{nutoomega}).

\subsection{Optimization quality}

It may be interesting to verify if larger or smaller recycling
probabilities lead to better optimization of the antibody
types. The statement that recycling may be necessary
to achieve a considerable affinity enhancement in a multi-step 
mutation process is widely discussed (Oprea et al., 2000)
and enforced by the present work. We would like to point
out that the relative number of {\it good} outcoming plasma
or memory cells increases with larger recycling probabilities
(see \abb{resqrun}).
\begin{figure}[ht]
\center{\hspace*{0mm} \epsfxsize=10cm \epsfysize10cm
            \epsfbox{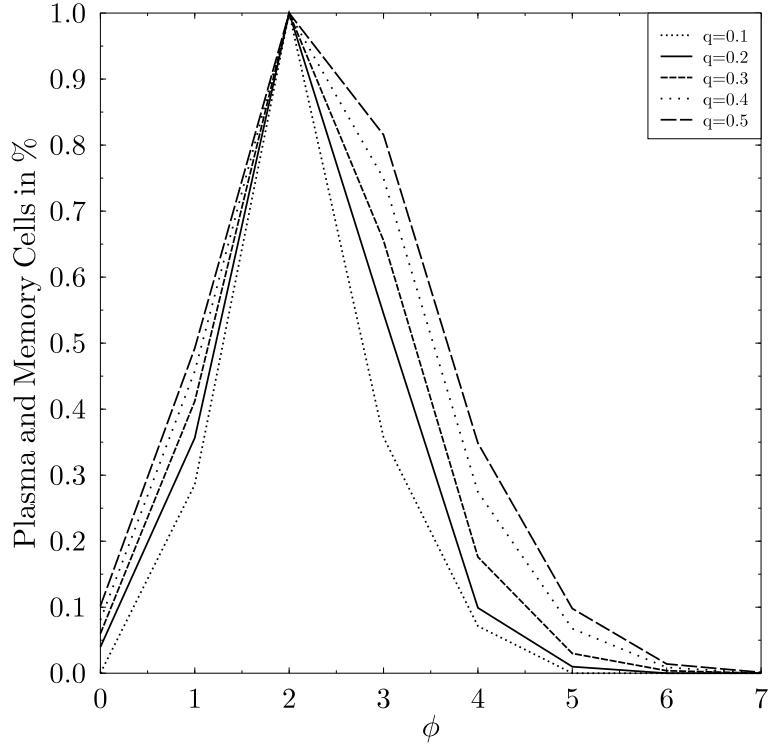}}
\caption[]{\sf Comparison of the total plasma and memory cell
distribution at day $21$ after immunization
on a cut along one of the coordinates in the shape space 
with the antigen at coordinate position $2$.
The parameters are uniquely determined by the 
experimental data with the exception of the remaining number of cells at
the end of the GC reaction, which is undetermined to allow different
recycling probabilities $1-q$. The distribution of the output cells
becomes sharper for larger $1-q$.}
\label{resqrun}
\end{figure}
In the shape space this corresponds
to a sharper peak of the distribution $\int_{0\,d}^{21\,d} dt\,O(\phi,t)$.
A too small recycling probability leads to a spreading of the
outcoming plasma and memory cells in the shape space, 
i.e.\,to a weaker specificity of the GC output.

On the other hand, we have already seen that a larger recycling probability
lowers the absolute number of produced plasma and memory cells
such that we are confronted with two competitive tendencies. Large
recycling probabilities lead to specific but weak GC reactions, while
small probabilities lead to unspecific but intense GC reactions. One may
agree, that the value of $80\%$ calculated here and supported by
experimental data, is a good compromise between these two
extremes.

\subsection{Start of somatic hypermutation}

There is experimental evidence that somatic
hypermutation does not occur during the first phase 
of proliferation of centroblasts in the environment
of the FDCs 
(Han et al., 1995b; Jacob et al., 1993; 
McHeyzer-Williams et al., 1993; Pascual et al., 1994b).
However, we checked if our model allow somatic hypermutation
to start during the proliferation phase.
The primary reason for the retardation of the output 
production is to give time to the GC to develop sufficient
good centroblasts before the GC population is weakened 
through an additional output. Therefore, one may
in principle think of a mutation starting
long before the selection process. In this scenario
the output production would start simultaneously with
selection and one is led to a two phase
process only: A first one with proliferation and 
somatic hypermutation and a second one with additional 
selection and output production.
The centroblast-types would already be spread out in the shape
space when selection and output start. 
\begin{figure}[ht]
\center{\hspace*{0mm} \epsfxsize=10cm \epsfysize10cm
            \epsfbox{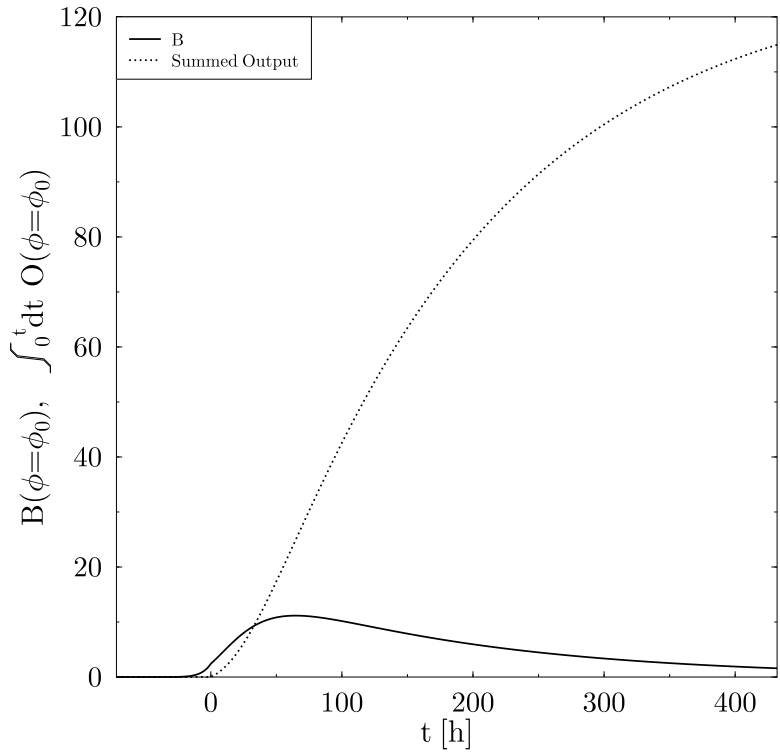}}
\caption[]{\sf The centroblast population and the 
integrated plasma/memory cell production at the position
of the antigen in the shape space. Except for the
time phases we used the same parameters as in \gl{bestpar}.
The centroblast population
shows a smooth enhancement due to somatic hypermutation
starting at $t=-48\,h$. The population growth is stopped by
the production of output cells starting at $t=0\,h$. The
slope of the summed output is substantially smaller than
expected from experiment. $21$ days after immunization
only $1$ cell remains. }
\label{boe61}
\end{figure}
There is no parameter set in accordance
with all experimental data. Especially, it is impossible
to get the correct output dynamics as described in
\gl{out126}. Typically we get $v_0=2.9$ being strongly
different from the required value and a much
smoother result (compare \abb{boe61}
to \abb{boe46}). In addition the distribution of plasma
and memory cells is not as well pronounced at the antigen 
position such that the GC reaction
gives rise to output cells with weaker specificity. Thus,
we deduce from our model that the dynamic of plasma 
and memory cell production in real GC reactions
does not allow somatic
hypermutation to occur considerably before the selection
process is started.

\section{Conclusions}

We developed a new model for the GC reaction
using its main functional elements. The GC reaction
is described by the evolution of a centroblast distribution
on an affinity shape space with respect to an initially
fixed antigen distribution.
On this shape space an affinity
function, modeling the complementarity of
antibody and antigen, was defined and its
functional behavior as well as its width were
deviated from affinity enhancements known
for real GC reactions.

In the model the reaction is decomposed into three phases
(see \abb{phases}). 
In the proliferation phase a few seeder cells multiply. 
Our model excludes somatic hypermutation 
to occur already at this stage of the GC development.
An optimization phase follows,
in which in addition to proliferation, mutation of the
antibody type and selection take place.
This leads to a competition of spreading and
peaking of the centroblast distribution in the
shape space.
In this phase, all positively selected cells reenter the
proliferation phase in the dark zone, i.e.\,no plasma
or memory cells are produced.
This occurs solely in the output production phase. In this
third phase all elements of the GC reaction are active.

The dynamical evolution of the cell population in
a GC is described by a set of coupled
linear differential equation. 
All parameters are determined
with experimental data and especially using
(Han et al., 1995b). The model shows the
typical behavior of GCs without any further
adjustments.
Due to the oligoclonal
character of GCs a large amount of identical 
centroblasts is produced in the proliferation phase.
These diffuse over the shape space in the second
phase, in which only those cells are retained that
exhibit a high affinity to the presented antigen.
As the {\it good} cells reenter the proliferation
process, this leads to a dominance of good cells
within $7$ to $8$ days after immunization in
accordance with experiments. The already
optimized cell distribution then gives rise to
plasma and memory cells leaving the FDC
environment in the output production phase, leading
to large numbers of output cells of optimized
type due to the long duration of this third phase.
The GC reaction is rather weak $21$ days after
immunization and only a small number of
cells remains in the environment of the FDCs.
This shows that the end of a GC reaction is
not necessarily coupled with an antigen consumption
or additional signals, but may occur simply due
to the (unchanged) dynamical development of the
GC.

Our main result is that the experimental data and
especially the evaluation of (Han et al., 1995b) lead
to the following view of the GC reaction.
The optimization phase lasts for not less than
$42\,h$ and not longer than $55\,h$. During this
time {\it all} selected centrocytes are recycled
and reenter the proliferation phase.
Without such a {\it non-output
phase} the number of optimal cells with respect
to the presented antigen is not large enough to
allow a considerable output rate during the
remaining life time of the GC. 
We looked for an experimental evidence for this
delayed production of output cells. Indeed, comparing
two experiments 
(Jacob et al., 1993; Pascual et al., 1994a) we found that 
plasma and memory cells were first observed more than 
two days later than mutated cells during a GC reaction. 
This confirms experimentally the existence of the 
non-output phase that we found in the model.
As output cells are not produced from the beginning
of the GC reaction, the output should be triggered by a special
signal, related to the affinity of the B-cells and
the corresponding antigen. But even in the
output production phase, the recycling probability
turns out to lie between $70\%$ and $85\%$
and thus is substantially larger than values
expected until now (Kesmir \& de Boer, 1999). 
We want to emphasize that such large recycling 
probabilities turned out to be crucial for a GC 
outcome of high specificity. Larger recycling probabilities
lead to specific GC reactions of low intensity, while
smaller ones to less specific but more intense 
GC reactions. The value of about $80\%$ that we
found represents a good compromise between these
competitive tendencies. On the other hand it is
not surprising that these recycled cells have not been
found experimentally: The absolute number of recycled
cells is very small compared to the total number
of cells present in the GC, so that a measured signal
for recycled cells is suppressed by more than two
orders of magnitudes. A better signal should be
observed in later stages of the GC reaction when
the cells of high affinity to the antigen are
already dominating.

Furthermore, we conclude that the whole selection
process including differentiation of centroblasts
to antibody presenting centrocytes, the multi-step
selection process itself and finally the recycling
take around $3$ or $4\,h$. This result is compatible
with the experimental observation that the inhibition
of apoptosis during the selection may take about
$1$ or $2\,h$.

It is interesting, that the produced cells
possess a certain broadness in the shape space, 
i.e.\,not only the centrocytes of optimal affinity to the antigen
differentiate to plasma and memory cells.
This may be of relevance for the resistance of an
immune system against a second immunization
with a mutated antigen. A considerable number
of memory cells with high affinity to the mutated
antigen may still be present for this reason.

The results of our model suggest new experiments.
A systematic and quantitative analysis of the GC
reaction in dependence of the initial conditions, 
i.e.\,the number of necessary point mutations to reach
the cell type of optimal affinity to the antigen, would
lead to new insights into the interdependence of the
essential elements of the GC. This may yield
hints about self-regulating processes.
Furthermore, it would be interesting to
analyze quantitatively
the total number of cells in the course
of the GC reaction.
Finally, it may be interesting to compare our model
with its continuous counterpart. 
The translation of \gl{modell} into continuous
space leads to a differential equation of Schr\"odinger
type with a Gaussian potential. 


\vfill
\eject
\newcounter{fig}
\section*{References}
\begin{list}{\normalsize
{\rm }}
{\usecounter{fig}
\setcounter{fig}{0}
\labelwidth0mm
\leftmargin8mm
\rightmargin0mm
\labelsep0mm
\topsep0mm
\parsep0mm
\itemsep0mm}
\itemindent-8mm
\normalsize

\item
{\sc Agarwal, A., Nayak, B.P. \& Rao, K.V.S.}
{\rm (1998)}.
{\rm B-Cell Responses to a Peptide Epitope -- VII -- Antigen-Dependent 
Modulation of the Germinal Center Reaction},
{\it J. Immunol.\/}
{\bf 161},
{\rm 5832--5841}.

\item
{\sc Berek, C. \& Milstein, C.}
{\rm (1987)}.
{\rm Mutation drift and repertoire shift in the maturation 
of the immune response},
{\it Immunol. Rev.\/}
{\bf 96},
{\rm 23--41}.

\item
{\sc Berek, C. \& Milstein, C.}
{\rm (1988)}.
{\rm The Dynamic Nature of the Antibody Repertoire},
{\it Immunol. Rev.\/}
{\bf 105},
{\rm 5--26}.

\item
{\sc Camacho, S.A., Koscovilbois, M.H. \& Berek, C.}
{\rm (1998)}.
{\rm The Dynamic Structure of the Germinal Center},
{\it Immunol. Today\/}
{\bf 19},
{\rm 511--514}.

\item
{\sc Cascalho, M., Ma, A., Lee, S., Masat, L. \& Wabl, M. }
{\rm (1996)}.
{\rm A Quasi-Monoclonal Mouse},
{\it Science\/}
{\bf 272},
{\rm 1649--1652}.

\item
{\sc Choe, J. \& Choi, Y.S.}
{\rm (1998)}.
{\rm IL-10 Interrupts Memory B-Cell Expansion in the Germinal 
Center by Inducing Differentiation into Plasma-Cells},
{\it Eur. J. Immunol.\/}
{\bf 28},
{\rm 508--515}.

\item
{\sc Cohen, J.J., Duke, R.C., Fadok, V.A. \& Sellins, K.S.}
{\rm (1992)}.
{\rm Apoptosis and programmed cell death in immunity},
{\it Annu. Rev. Immunol.\/}
{\bf 10},
{\rm 267--293}.

\item
{\sc Dubois, B., Barth\'el\'emy, C., Durand, I., Liu, Y.-J., Caux, 
C. \& Bri\`ere, F.}
{\rm (1999)}.
{\rm Toward a Role of Dendritic Cells in the Germinal Center 
Reaction -- Triggering of B-Cell Proliferation and Isotype 
Switching},
{\it J. Immunol.\/}
{\bf 162},
{\rm 3428--3436}.

\item
{\sc Eijk, M. van \& Groot, C. de},
{\rm (1999)}.
{\rm Germinal Center B-Cell Apoptosis Requires Both Caspase 
and Cathepsin Activity},
{\it J. Immunol.\/}
{\bf 163},
{\rm 2478--2482}.

\item
{\sc Fischer, M.B., Goerg, S., Shen, L.M., Prodeus, A.P., Goodnow, 
C.C., Kelsoe, G. \& Carroll, M.C.}
{\rm (1998)}.
{\rm Dependence of Germinal Center B-Cells on Expression 
of Cd21/Cd35 for Survival},
{\it Science\/}
{\bf 280},
{\rm 582--585}.

\item
{\sc Fliedner, T.M.}
{\rm (1967)}.
{\rm On the origin of tingible bodies in germinal centers 
in immune responses},
{\rm in: H. Cottier (Hrsg.)},
{\rm Germinal Centers in Immune Responses}.
{\rm Springer},
{\rm Berlin},
{\rm pp.\,218--224}.

\item
{\sc Grouard, G., De Bouteiller, O., Banchereau, J. \& Liu, Y.-J.}
{\rm (1995)}.
{\rm Human follicular dendritic cells enhance cytokine-dependent 
growth and differentiation of CD40-activated B cells},
{\it J. Immunol.\/}
{\bf 155},
{\rm 3345--3352}.

\item
{\sc Han, S.H., Hathcock, K., Zheng, B., Kelper, T.B., Hodes, 
R. \& Kelsoe, G.}
{\rm (1995a)}.
{\rm Cellular Interaction in Germinal Centers: Roles of 
CD40-Ligand and B7-1 and B7-2 in Established Germinal 
Centers},
{\it J. Immunol.\/}
{\bf 155},
{\rm 556--567}.

\item
{\sc Han, S.H., Zheng, B., Dal Porto, J. \& Kelsoe, G.}
{\rm (1995b)}.
{\rm In situ Studies of the Primary Immune Response to 
(4--Hydroxy--3--Nitrophenyl) 
Acetyl IV. Affinity-Dependent, Antigen-Driven B-Cell 
Apoptosis in Germinal Centers as a Mechanism for Maintaining 
Self-Tolerance},
{\it J. Exp. Med.\/}
{\bf 182},
{\rm 1635--1644}.

\item
{\sc Hanna, M.G.}
{\rm (1964)}.
{\rm An autoradiographic study of the germinal center in 
spleen white pulp during early intervals of the immune 
response},
{\it Lab. Invest.\/}
{\bf 13},
{\rm 95--104}.

\item
{\sc Hardie, D.L., Johnson, G.D. \& MacLennan, I.C.M.}
{\rm (1993)}.
{\rm Quantitative analysis of molecules which distinguish 
functional compartments in germinal centers},
{\it Eur. J. Immunol.\/}
{\bf 23},
{\rm 997--1004}.

\item
{\sc Jacob, J., Kassir, R. \& Kelsoe, G.}
{\rm (1991)}.
{\rm In situ studies of the primary immune response 
to (4-hydroxy-3-nitrophenyl)acetyl. 
I. The architecture and dynamics of responding cell 
populations},
{\it J. Exp. Med.\/}
{\bf 173},
{\rm 1165--1175}.

\item
{\sc Jacob, J., Przylepa, J., Miller, C. \& Kelsoe, G.}
{\rm (1993)}.
{\rm In situ studies of the primary response 
to (4-hydroxy-3-nitrophenyl)acetyl. 
III. The kinetics of V region mutation and selection 
in germinal center B cells},
{\it J. Exp. Med.\/}
{\bf 178},
{\rm 1293--1307}.

\item
{\sc Janeway, C.A. \& Travers, P.}
{\rm (1997)}.
{\rm Immunologie}.
{\rm Spektrum Akademischer Verlag},
{\rm Heidelberg, Berlin, Oxford}.

\item
{\sc Kelsoe, G.}
{\rm (1996)}.
{\rm The germinal center: a crucible for lymphocyte selection},
{\it Semin. Immunol.\/}
{\bf 8},
{\rm 179--184}.

\item
{\sc Kepler, T.B. \& Perelson, A.S.}
{\rm (1993)}.
{\rm Cyclic re-entry of germinal center B cells and the 
efficiency of affinity maturation},
{\it Immunol. Today\/}
{\bf 14},
{\rm 412--415}.

\item
{\sc Kesmir, C. \& Boer, R.J.\,de},
{\rm (1999)}.
{\rm A Mathematical Model on Germinal Center Kinetics and 
Termination},
{\it J. Immunol.\/}
{\bf 163},
{\rm 2463--2469}.

\item
{\sc Kroese, F.G., Wubbena, A.S., Seijen, H.G. \& Nieuwenhuis, 
P.}
{\rm (1987)}.
{\rm Germinal centers develop oligoclonally},
{\it Eur. J. Immunol.\/}
{\bf 17},
{\rm 1069--1072}.

\item
{\sc K\"uppers, R., Zhao, M., Hansmann, M.L. \& Rajewsky, K.}
{\rm (1993)}.
{\rm Tracing B Cell Development in Human Germinal Centers 
by Molecular Analysis of Single Cells Picked from Histological 
Sections},
{\it EMBO J.\/}
{\bf 12},
{\rm 4955--4967}.

\item
{\sc Lindhout, E., Lakeman, A. \& Groot, C. de},
{\rm (1995)}.
{\rm Follicular dendritic cells inhibit apoptosis in human 
B lymphocytes by rapid and irreversible blockade of 
preexisting endonuclease},
{\it J. Exp. Med.\/}
{\bf 181},
{\rm (1985--1995}.

\item
{\sc Lindhout, E., Mevissen, M.L., Kwekkeboom, J., Tager, J.M. \& Groot, 
C. de},
{\rm (1993)}.
{\rm Direct evidence that human follicular dendritic cells 
(FDC) rescue germinal centre B cells from death by 
apoptosis},
{\it Clin. Exp. Immunol.\/}
{\bf 91},
{\rm 330--336}.

\item
{\sc Lindhout, E., Koopman, G., Pals, S.T. \& Groot, 
C.\,de},
{\rm (1997)}.
{\rm Triple check for antigen specificity of B cells during 
germinal centre reactions},
{\it Immunol. Today\/}
{\bf 18},
{\rm 573--576}.

\item
{\sc Liu, Y.-J., Barth\'el\'emy, C., De Bouteiller, O. \& Banchereau, 
J.}
{\rm (1994)}.
{\rm The differences in survival and phenotype between centroblasts 
and centrocytes},
{\it Adv. Exp. Med. Biol.\/}
{\bf 355},
{\rm 213--218}.

\item
{\sc Liu, Y.-J., Joshua, D.E., Williams, G.T., Smith, C.A., Gordon, 
J. \& MacLennan, I.C.}
{\rm (1989)}.
{\rm Mechanism of antigen-driven selection in germinal centres},
{\it Nature\/}
{\bf 342},
{\rm 929--931}.

\item
{\sc Liu, Y.-J., Zhang, J., Lane, P.J., Chan, E.Y. \& MacLennan, 
I.C.M.}
{\rm (1991)}.
{\rm Sites of specific B cell activation in primary and 
secondary responses to T cell-dependent and T cell-independent 
antigens},
{\it Eur. J. Immunol.\/}
{\bf 21},
{\rm 2951--2962}.

\item
{\sc MacLennan, I.C.M.}
{\rm (1994)}.
{\rm Germinal Centers},
{\it Annu. Rev. Immunol.\/}
{\bf 12},
{\rm 117--139}.

\item
{\sc McHeyzer-Williams, M.G., McLean, M.J., Labor, P.A. \& Nossal, 
G.V.J.}
{\rm (1993)}.
{\rm Antigen-driven B cell differentiation in vivo},
{\it J. Exp. Med.\/}
{\bf 178},
{\rm 295--307}.

\item
{\sc Nossal, G.}
{\rm (1991)}.
{\rm The molecular and cellular basis of affinity maturation 
in the antibody response},
{\it Cell\/}
{\bf 68},
{\rm 1--2}.

\item
{\sc Oprea, M. \& Perelson, A.S.}
{\rm (1996)}.
{\rm Exploring the Mechanism of Primary Antibody Responses 
to T-Cell-Dependent Antigen},
{\it J. Theor. Biol.\/}
{\bf 181},
{\rm 215--236}.

\item
{\sc Oprea, M. \& Perelson, A.S.}
{\rm (1997)}.
{\rm Somatic mutation leads to efficient affinity maturation 
when centrocytes recycle back to centroblasts},
{\it J. Immunol.\/}
{\bf 158},
{\rm 5155--5162}.

\item
{\sc Oprea, M., Nimwegen, E.\,van \& Perelson, A.S.}
{\rm 2000)}.
{\rm Dynamics of One-pass Germinal Center Models: Implications 
for Affinity Maturation},
{\it Bull. Math. Biol.\/}
{\bf 62},
{\rm 121--153}.

\item
{\sc Pascual, V., Cha, S., Gershwin, M.E., Capra, J.D. \& Leung, 
P.S.C.}
{\rm (1994a)}.
{\rm Nucleotide Sequence Analysis of Natural and Combinatorial 
Anti-PDC-E2 Antibodies in Patients with Primary Biliary 
Cirrhosis},
{\it J. Immunol.\/}
{\bf 152},
{\rm 2577--2585}.

\item
{\sc Pascual, V., Liu, Y.-J., Magalski, A., De Bouteiller, 
O., Banchereau, J. \& Capra, J.D.}
{\rm (1994b)}.
{\rm Analysis of somatic mutation in five B cell subsets 
of human tonsil},
{\it J. Exp. Med.\/}
{\bf 180},
{\rm 329--339}.

\item
{\sc Perelson, A.S. \& Oster, G.F.}
{\rm (1979)}.
{\rm Theoretical Studies of Clonal Selection: Minimal Antibody 
Repertoire Size and Reliability of Self-Non-self Discrimination},
{\it J. Theor. Biol.\/}
{\bf 81},
{\rm 645--670}.

\item
{\sc Perelson, A.S. \& Wiegel, F.W.}
{\rm (1999)}.
{\rm Some Design Principles for Immune System Recognition},
{\it Complexity\/}
{\bf 4},
{\rm 29--37}.

\item
{\sc Radmacher, M.D., Kelsoe, G. \& Kepler, T.B.}
{\rm (1998)}.
{\rm Predicted and Inferred Waiting-Times for Key Mutations 
in the Germinal Center Reaction -- Evidence for Stochasticity 
in Selection},
{\it Immunol. Cell Biol.\/}
{\bf 76},
{\rm 373--381}.

\item
{\sc Rundell, A., Decarlo, R., Hogenesch, H. \& Doerschuk, P.}
{\rm (1998)}.
{\rm The Humoral Immune-Response to Haemophilus-Influenzae 
Type-B -- A Mathematical-Model Based on T-Zone and Germinal 
Center B-Cell Dynamics},
{\it J. Theor. Biol.\/}
{\bf 194},
{\rm 341--381}.

\item
{\sc Smith, K., Light, A., Nossal, G. \& Tarlington, D.}
{\rm (1997)}.
{\rm The extent of affinity maturation differs between the 
memory and antibody-forming cell compartments in the 
primary immune response},
{\it EMBO J.\/}
{\bf 16},
{\rm 2996--3006}.

\item
{\sc Tew, J. \& Mandel, T.}
{\rm (1979)}.
{\rm Prolonged antigen half-life in the lymphoid follicles 
of specifically immunized mice},
{\it Immunology\/}
{\bf 37},
{\rm 69--76}.

\item
{\sc de Vinuesa, C.G., Cook, M.C., Ball, J., Drew, M., Sunners, Y.,
Cascalho, M., Wabl, M., Klaus, G.G.B. \& MacLennan, C.M.}
{\rm (2000)}.
{\rm Germinal Centers without T Cells},
{\it J. Exp. Med.\/}
{\bf 191},
{\rm 485--493}.

\item
{\sc Wedemayer, G.J., Patten, P.A., Wang, L.H., Schultz, 
P.G. \& Stevens, R.C.}
{\rm (1997)}.
{\rm Structural insights into the evolution of an antibody 
combining site},
{\it Science\/}
{\bf 276},
{\rm 1665--1669}.

\item
{\sc Weigert, M., Cesari, I., Yonkovitch, S. \& Cohn, M.}
{\rm (1970)}.
{\rm Variability in the light chain sequences of mouse antibody},
{\it Nature\/}
{\bf 228},
{\rm 1045--1047}.

\item
{\sc Zhang, J., MacLennan, I.C.M., Liu, Y.-J. \& Land, P.J.L.}
{\rm (1988)}.
{\rm Is rapid proliferation in B centroblasts linked to 
somatic mutation in memory B cell clones},
{\it Immunol. Lett.\/}
{\bf 18},
{\rm 297--299}.

\end{list}

\end{document}